# Self-Assembled Nanowires with Giant Rashba-Type Band Splitting

Jewook Park[1], Sung Won Jung[1], Min-Cherl Jung[1], Hiroyuki Yamane[2], Nobuhiro Kosugi[2], and Han Woong Yeom[1*]

[1]Department of Physics and Center for Low Dimensional Electronic Symmetry,
Pohang University of Science and Technology, Pohang 790-784, Korea
[2]Department of Photo-Molecular Science, Institute for Molecular Science, Okazaki 444-8585, Japan

We investigated Pt-induced nanowires on the Si(110) surface using scanning tunneling microscopy (STM) and angle-resolved photoemission (ARP). High resolution STM images show a well-ordered nanowire array of 1.6 nm width and 2.7 nm separation. ARP reveals fully occupied one dimensional (1D) bands with a Rashba-type split dispersion. Local $dI/dV$ spectra further indicate well confined 1D electron channels on the nanowires, whose density of states characteristics are consistent with the Rashba-type band splitting. The observed energy and momentum splitting of the bands are among the largest ever reported for Rashba systems, suggesting the Pt-Si nanowire as a unique 1D giant Rashba system. This self-assembled nanowire can be exploited for silicon-based spintronics devices as well as the quest for Majorana Fermions.

DOI:                         PACS umber(s): 68.37.Ef, 73.20.At, 79.60.Bm, 71.70.Ej

The Rashba effect [1], a spin splitting in surface or interface electronic bands due to the spin-orbit coupling (SOC) and the broken inversion symmetry, was recently brought to attention mainly because of the possibility of spintronics without magnetic field [2-4]. Although a larger band splitting is desirable for such applications, the splitting is usually not sufficiently large in most Rashba systems, such as surface states of metal (Au, Bi, and Sb) crystals [5-7], semiconductor heterostructures (InGaAs/AlGaAs, Si/SiGe) [8, 9], metal quantum wells (Au/W(110)) [10], and vertical quantum dots ($In_{0.05}Ga_{0.95}As$) [11]. However, it was recently found that the Rashba splitting can be greatly enhanced in a few materials systems, such as the Bi/Ag surface alloy [12], the Bi monolayer on Si [13], the Bi quantum films on Cu(111) [14], the bulk BiTeI [15], and the Ir(111) surface [16]. These systems exhibit at least an order of magnitude larger spin splitting and are categorized as 'giant Rashba systems'.

Although most Rashba systems exist in two-dimensional surfaces and films, one-dimensional (1D) nanowires with a strong SOC have obvious merit in manipulating spin carriers [17]. Moreover, recent theoretical studies suggest that they can be useful for investigating helical liquid states [18-22] and realizing exotic particles of Majorana Fermions [23, 24]. Thus far, the 1D systems, such as carbon nanotubes [25] and semiconductor quantum wires [26], exhibit only small spin splittings. More recently, a substantial Rashba effect was observed for self-assembled Au nanowires on vicinal Si surfaces [27-29]. Even so, the strength of SOC for these systems is far from the giant Rashba effect regime.

In this Letter, we introduce a new 1D system with the giant Rashba effect; the Pt-induced nanowire on the Si(110) surface (hereafter, the Pt-Si nanowire). We characterized the structure of the Pt-Si nanowire and revealed its 1D electronic states using scanning tunneling microscopy (STM) and spectroscopy (STS). Angle-resolved photoemission (ARP) measurements disclosed the 1D Rashba-type bands with a giant splitting of band dispersions and their gapless crossing in both momentum and energy. The band dispersions and the density of states in ARP and STS are fully consistent with the Rashba-type splitting. Thus, we suggest the Pt-Si nanowire as a unique 1D giant Rashba system.

The clean As-doped Si(110) ($\rho \sim 0.01$ Ωcm) substrate was prepared by 873 K degassing and 1523 K flash heating in ultra-high vacuum. The Si(110) surface could act as a template for the nanowire growth due to its intrinsically anisotropic atomic structure [30, 31]. A well-ordered nanowire array was formed on this substrate after a nominal one monolayer deposition of Pt at 1023 K and 873 K annealing. We confirmed that it corresponds to the previously reported "6"×5 phase with

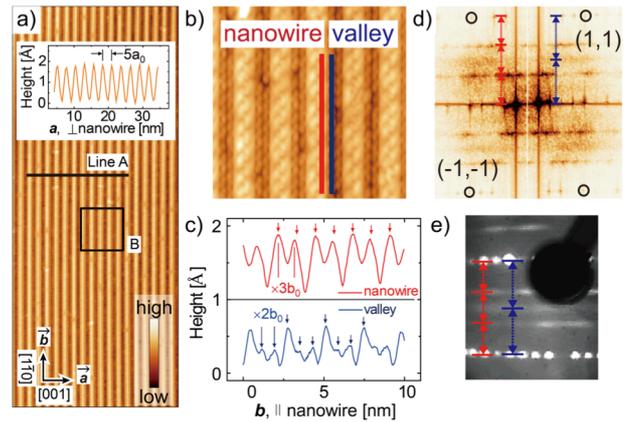

FIG. 1 (a) STM image of Pt-Si nanowires (120 nm × 50 nm at 1.0 V). Inset is the line profile along the line A indicating the width (1.6 nm) and separation ($5a_0$) of nanowires. (b) Magnified STM image of the region B in (a) (13.7 nm × 13.7 nm at 500 mV). (c) Line profiles on the nanowire and valley along the red and blue lines in (b), which show $3b_0$ and $2b_0$ periodicities, respectively. (d) Fourier transformed STM image and (e) low-energy-electron diffraction pattern display a mixture of $3b_0$ and $2b_0$ periodicities along the $b$ axis.

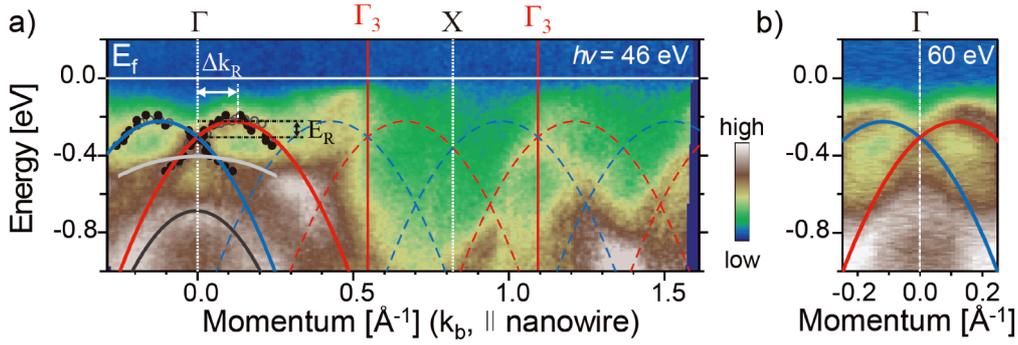

FIG. 2 (a) ARP intensity plot of Pt-Si nanowires along the nanowire direction, the $b$ axis ($h\nu = 46$ eV). Red and blue parabolic lines around the $\Gamma$ point are interpreted as Rashba-type spin-split bands. The spectral peak positions from the momentum (filled circles) and energy (open circles) distribution curves were used in the parabolic fitting of band dispersions. These Rashba-type bands are repeated along nanowires with a periodicity of $2\pi/3b_0$. $\Gamma$ and X points indicate the Si(110) 1×1 brillouin zone center and boundary, respectively, and $\Gamma_3$ is the ×3 surface brillouin zone center. $E_R$ and $\Delta k_R$ are 81 meV and 0.12 Å$^{-1}$, respectively. The black and gray parabolic lines around the $\Gamma$ point are the heavy hole band and the valence band edge of Si(110) bulk, respectively. (b) Similar ARP intensity plot around the $\Gamma$ point taken at a different photon energy ($h\nu = 60$ eV).

a 1D structure along the [1$\bar{1}$0] direction, using low-energy-electron diffraction [Fig. 1(e)] [32].

The STM and STS experiments were carried out using a commercial cryogenic STM (Unisoku, Japan) at 78 K. Figures 1(a) and 1(b) show STM images of the well-ordered Pt-Si nanowires. The nanowires have a width of 1.6 nm and are regularly-arrayed with a period of 2.7 nm [$5a_0$, $a_0 = 0.543$ nm and $b_0 = 0.384$ nm for the ideal Si(110) surface unit cell in the [001] and [1$\bar{1}$0] directions, respectively]. The enlarged STM image of Fig. 1(b) shows bright and dark stripes, which we call 'nanowires' and 'valleys', respectively. Based on the line profiles [Fig. 1(c)], we were able to precisely determine the local periodicities along the wires and valleys. The protrusions along nanowires and valleys have primary $3b_0$ and $2b_0$ periodicities, respectively, which are commonly modulated into a longer secondary periodicity of $6b_0$. The $3b_0$ and $2b_0$ periodicities are also reflected in the Fourier transformed image [Fig. 1(d)] and the electron diffraction pattern [Fig. 1(e)].

The bands structure of these nanowires were revealed by low temperature (~15 K) ARP measurements at an undulator beam line BL6U [33] of UVSOR synchrotron radiation facility (Okazaki, Japan) using photon energy ($h\nu$) of 46 ~ 80 eV and a standard hemispherical analyzer. The overall energy and momentum resolutions were better than 20 meV and 0.03 Å$^{-1}$, respectively. Similar ARP measurements were also conducted in a home-built system at 100 K using a high-flux He discharge lamp ($h\nu$ = 21.2 eV). The overall energy and momentum resolutions were better than 20 meV and 0.02 Å$^{-1}$, respectively.

Figure 2 shows electronic band dispersions along the nanowires (the $b$ axis) measured at two different photon energies. We observed two parabolic bands (blue and red lines) along the nanowires as well as the well-known Si bulk bands (black and gray parabolas) around the $\Gamma$ point. The two characteristic parabolic bands, whose band top is located at 0.2 eV below the Fermi level, are obviously surface state bands because they are largely located within the bulk band gap [31, 34] and its dispersion does not change with respect to the incident photon energy [Fig. 2(b)] [35]. Their dispersions are carefully quantified by Lorentzian fits of the energy and momentum distribution curves as plotted in Fig. 2(a) [35]. Although the intensity is largely modulated, one can trace the bands following the $2\pi/3b_0$ periodicity (dashed parabolas). The quasi-1D character of the parabolic bands is also confirmed; their dispersion normal to the nanowires is less than 150 meV [35]. Therefore, we can conclude that these bands are localized along the nanowire parts of the surface, *i. e.*, the bright chains in STM images with a $3b_0$ periodicity. This conclusion is further supported by the STS measurements as discussed below. What is very important is that the two parabolic bands with exactly the same effective mass and binding energy cross each other gaplessly at $\Gamma$ [35]. This kind of dispersion cannot easily be explained without invoking the Rashba-type SOC.

We also obtained detailed spatially resolved $dI/dV$ spectra for the nanowires and valleys by STM and STS. From the high resolution topographic image [Fig. 3(a)],

TABLE. 1 Selected Rashba systems and parameters characterizing the spin splitting of their bands; Rashba energy ($E_R$), momentum offset ($\Delta k_R$), and Rashba parameter ($\alpha_R$).

| Materials | $E_R$ [meV] | $\Delta k_R$ [Å$^{-1}$] | $\alpha_R$ [eVÅ] | Ref. |
|---|---|---|---|---|
| Bulk BiTeI (3D) | 100 | 0.052 | 3.8 | [15] |
| Bi/Ag(111) surface alloy (2D) | 200 | 0.13 | 3.05 | [12] |
| Pt-Si nanowire (1D) | 81 | 0.12 | 1.36 | This work |
| Au/W(110) quantum well (2D) | 0.2 | 0.0025 | 0.16 | [10] |
| Au(111) surface state (2D) | 2.1 | 0.012 | 0.33 | [5] |
| InGaAs/InAlAs (1D) | <1 | 0.028 | 0.07 | [8] |
| Si(557)-Au nanowire (1D) | N/A | 0.05 | N/A | [29] |
| Vertical In$_{0.05}$Ga$_{0.95}$As/GaAs quantum dot (0D) | 0.12 | 0.002 ~ 0.003 | 0.08 ~ 0.12 | [11] |

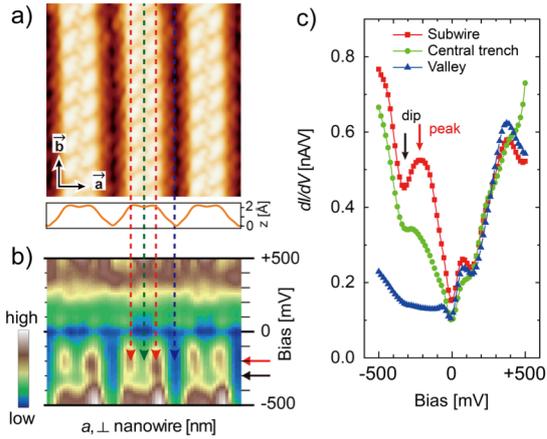
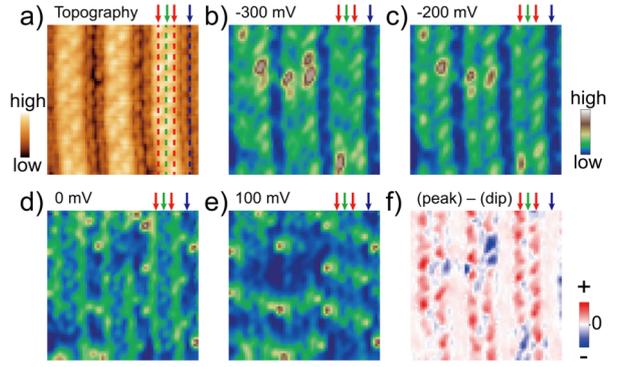

FIG. 3 (a) Detailed filled-state STM image of the Pt-Si nanowires (8.3 nm × 8.3 nm at -500 mV). The boxed curve at bottom is the averaged height across nanowires. (b) Line-averaged (along nanowires) dI/dV spectra across nanowires for the same area as (a). dI/dV conductance curves are obtained on 64 × 64 grid points over the area of (a) and the intensity of the dI/dV curves are converted to the color scale. Dotted lines with arrows indicate the positions of subwires (red), a central trench (green), and a valley (blue). (c) dI/dV curves averaged along a subwire (red), a central trench (green), and a valley (blue). The small arrows in (b) and (c) indicate the characteristic peak-and-dip feature in LDOS. During the dI/dV conductance measurement the bias voltage was set at +500 mV.

FIG. 4 Simultaneously acquired (a) topography (8.3 nm × 8.3 nm at 500 mV) and (b)-(e) bias dependent dI/dV conductance maps. Bias voltages are denoted on the maps. Red, green, and blue arrows and the dashed lines indicate subwires, central trenches, and valleys, respectively, as in Fig 3. (f) Subtraction of the spectroscopy map at -300 mV from that at -200 mV. This corresponds to the difference of the dI/dV conductance of the peak and the dip.

we identified that each nanowire has two identical 'subwires' (red dotted lines) and a 'central trench' (green dotted line). We took dI/dV spectra for a 64×64 grid over the surface area shown in this image. Figure 3(b) shows the line-averaged spectra along the b axis, plotted along the a axis crossing the nanowires. The dI/dV curves unveil distinct features on subwires, centers trenches, and valleys, especially for filled states. In particular, the STS spectra show a strong local density of states (LDOS) peak and a characteristic dip at -200 and -300 mV, respectively, which are well confined on the nanowires (subwires). The peak and dip energies quantitatively agree with the band dispersions observed by ARP in Fig. 2. The peak at -200 mV corresponds to the LDOS singularity of the Rashba-type surface band dispersions, which is expected at the top of the parabolic bands [36]. The characteristic dip at -300 mV corresponds to the Dirac-like crossing of the bands as observed in graphene and topological surface states [37].

The lateral maps of the dI/dV spectra shown in Fig. 4 further confirm that the spectral features at -200 and -300 mV are well confined on the nanowires (the subwire parts) and extend along them. Apparent positive value (red) on the subwires in Fig. 4(f) represent that the peak-and-dip feature, which originate from the Rashba effect, is confined on the subwires. The confined spectral feature is in clear contrast to, for example, the empty state spectral peak at +100 mV, which is localized on certain atomic sites in valleys [35]. Summarizing the ARP and STS results, we conclude that the Pt-Si nanowires host fully occupied 1D Rashba-type spin split bands. The fingerprint of the Rashba splitting in STS, namely the peak-dip structure does not vary significantly from point to point along the top of the subwires [33]. That is, the Rashba splitting is robustly connected to the Pt-Si wire structure. Since the wire structure itself is uniformly formed over the surface with the wire length up to the micron scale, the current Rashba effect is not just a local effect as also observed by spatially averaging technique of ARP.

The most exciting finding for the Pt-Si nanowire system is the size of the Rashba-type band splitting as well as its 1D character. The Rashba Hamiltonian is given as, $H_R = \alpha_R \sigma(k_\parallel \times e_z)$, where the coupling constant $\alpha_R$ is the Rashba parameter, $\sigma$ the Pauli matrix, $k_\parallel$ the in-plane momentum, and $e_z$ the surface normal vector. Due to SOC, the resulting dispersion of the spin-split bands are described by $E^\pm(k) = (\hbar^2 k_\parallel^2/2m^*) \pm \alpha_R |k_\parallel|$, where $m^*$ represents the electron effective mass. Tab. 1 shows the reported band parameters ($E_R$, $\Delta k_R$, and $\alpha_R$) of various Rashba systems, where the Rashba energy ($E_R$) is the binding energy difference between the crossing point and the band top (edge), $\Delta k_R$ the momentum splitting of the Rashba bands from the crossing point, and the Rashba parameter $\alpha_R = \hbar^2 \Delta k_R/m^*$. In the present ARP data, $E_R$ and $\Delta k_R$ are 81 meV and 0.12 Å$^{-1}$, respectively. Using the fitted effective mass ($m^*$ = -0.66 $m_0$) and the measured $\Delta k_R$, we could deduce $\alpha_R$ = 1.36 eVÅ. The observed $E_R$ and $\Delta k_R$ are among the largest ever reported for Rashba systems and belongs to very few giant Rashba cases (Tab.1) [12-16]. We thus suggest that this system is the first 1D giant Rashba system ever discovered. As also compared in Tab.1, the present Rashba effect is an order of magnitude larger than the conventional 1D Rashba system based on semiconductor quantum wire structures [8]. This effect is also three or four times larger than the recent observation of Au wires on vicinal Si surfaces [26].

According to the present understanding, the strength of the spin splitting in a Rashba system is associated with both the atomic SOC and the potential gradient in the surface normal direction [38-40]. In addition, a recent study reported that the local angular momentum should be considered to understand the Rashba effect [41]. Several reports on giant Rashba systems further concluded that the in-plane potential gradient also plays a crucial role to enhance the Rashba spin splitting [12, 13]. In our $dI/dV$ conductance images [Figs. 4(b) and 4(c)], distinct 1D electronic channels are displayed along the nanowires, which implicate a strong in-plane potential gradient on the surface through the strong transverse modulation of the electron distribution in addition to the atomic corrugation. We suggest that the nanowire structure and the 1D electronic anisotropy are related to the giant Rashba effect. The larger Rashba splitting compared with Au/Si [26] are thought come from the $d$ electron contribution; the $d$ band of Pt is much closer to the Fermi energy than that of Au. However, the further microscopic understanding of its origin has to await the detailed information on the atomic structure of the nanowires. At present, we presume that the Pt-Si nanowires are anisotropic Pt silicide crystals as observed for rare earth metals on other Si surfaces [42]. From the epitaxial lattice matching and the growth temperature, PtSi(101)/Si(110) is a possible structural basis of the nanowires.

In summary, we investigated the electronic structure of Pt-Si nanowires on Si(110) with STM/STS and ARP. Along the regularly-arrayed nanowires, the clear 1D electronic channels were observed, whose band dispersions and density of states show characteristic features of the Rashba-type band splitting. The size of Rashba splitting is huge enough to suggest this system as the first 1D giant Rashba system, which would be important for spintronics applications and the quest for Majorana Fermions.

This work was supported by MOST of Korea through the Center for Low Dimensional Electronic Symmetry of the KRF CRi program. H. Y. and N. K. acknowledge a Grant-in-Aid for Young Scientists (A) (No. 24685032) and a Grant-in-Aid for Scientific Research (A) (No. 23245007) from JSPS of Japan .

*yeom@postech.ac.kr